\documentclass[reqno,12pt]{article}

\usepackage{graphicx}
\usepackage{amsmath,amsthm,amssymb}

\chardef\bslash=`\\ 

\newtheorem{thm}{Theorem}
\newtheorem{cor}[thm]{Corollary}
\newtheorem{lem}[thm]{Lemma}
\newtheorem{prop}[thm]{Proposition}

\theoremstyle{definition}

\theoremstyle{remark}

\newtheorem*{thm2}{Theorem}

\newcommand{\eval}[2][\right]{\relax
 \ifx#1\right\relax \left.\fi#2#1\rvert}

\begin{document}
\title{\bf{On the existence of self-similar spherically symmetric wave maps coupled to gravity}}

\author{Piotr Bizo\'n\footnotemark[1]{}\,\; and
 Arthur Wasserman\footnotemark[2]{}\\
 \footnotemark[1]{} \small{\textit{Institute of Physics,
  Jagellonian University, Krak\'ow, Poland}}\\
  \footnotemark[2]{} \small{\textit{Department of Mathematics,
  University of Michigan, Ann Arbor, Michigan}}}
\maketitle
\begin{abstract}
\noindent We present a detailed analytical study of spherically
symmetric self-similar solutions in the $SU(2)$ sigma model
coupled to gravity. Using a shooting argument we prove that there
is a countable family of  solutions which are analytic  inside the
past self-similarity horizon. In addition, we  show that for
sufficiently small values of the coupling constant these solutions
possess a regular future self-similarity horizon and thus are
examples of naked singularities. One of the solutions constructed
here has been recently found as the critical solution at the
threshold of black hole formation.
\end{abstract}

\section{Introduction}
In this paper we continue our investigations, started
in~\cite{bi_wa} (referred to as I), of wave maps coupled to
gravity, that is solutions of Einstein's equations with an $SU(2)$
sigma field as matter. We found numerically in I that for
$\alpha<1/2$ ($\alpha$ is the dimensionless coupling constant) the
model admits a countable family of continuously self-similar (CSS)
solutions, labeled by an integer nodal index $n=0, 1,\ldots$, that
are analytic inside the past light cone of the singularity. We
also provided evidence that the $n$th CSS solution can be
continued up to the future light cone of the singularity if
$\alpha<\alpha_n$, where $\{\alpha_n\}$ is an increasing sequence
bounded above by $1/2$. The purpose of this paper is to make the
results of I into theorem--proof rigorous mathematics. This is
accomplished by applying  a shooting argument to the resulting
dynamical system. We note that the case $\alpha=0$ was previously
analyzed in~\cite{bi}.

The physical importance of the CSS solutions considered here was
discussed in I, in particular we conjectured that in a certain
parameter range ($\alpha_0<\alpha<\alpha_1$) the $n=1$ solution is
a critical solution at the threshold of black hole formation. This
conjecture has been recently confirmed in numerical studies of the
critical behaviour~\cite{vienna} and in the linear stability
analysis~\cite{lechner}. As far as we know, this is the only case
where the existence of a self-similar solution, which was
numerically found as the critical solution in gravitational
collapse, has been established rigorously.
\section{Setup}
For the reader's convenience we repeat from I the basic setting
for the problem. Let $X: M \rightarrow N$ be a map from a
spacetime $(M,g_{ab})$ into a Riemannian manifold $(N,G_{AB})$.
Wave maps coupled to gravity are defined as extrema of the action
\begin{equation}
S = \int_M \left(\frac{R}{16 \pi G}  + L_{WM}\right) dv_g
\end{equation}
with the Lagrangian density
\begin{equation}
 L_{WM} = -\frac{f^2_{\pi}}{2} g^{ab} \partial_a X^A \partial_b X^B
 G_{AB}.
\end{equation}
Here $G$ is Newton's constant and $f^2_{\pi}$ is the wave map
coupling constant. The product $\alpha=4\pi G f^2_{\pi}$ is
dimensionless. The field equations derived from (1) are the wave
map equation
\begin{equation}
\square_g X^A + \Gamma_{BC}^A(X) \partial_aX^B \partial_bX^C
g^{ab}=0,
\end{equation}
where $\Gamma_{BC}^A(X)$ are the Christoffel symbols of the target
metric $G_{AB}$ and $\square_g$ is the d'Alembertian associated
with the metric $g_{ab}$, and  the Einstein equations
$R_{ab}-\frac{1}{2} g_{ab} R = 8 \pi G T_{ab}$ with the
stress-energy tensor
\begin{equation}
 T_{ab} = f^2_{\pi} \left(\partial_a X^A \partial_b X^B
 -\frac{1}{2} g_{ab}( g^{cd} \partial_c X^A \partial_d X^B)\right)
 G_{AB}.
\end{equation}
As a target
manifold we take the three-sphere $S^3$ with the standard
metric in polar coordinates $X^A=(F,\Theta,\Phi)$
\begin{equation}
G_{AB} dX^A dX^B = dF^2 + \sin^2{\!F} \:(d\Theta^2 +
\sin^2{\!\Theta}\: d\Phi^2).
\end{equation}
For the domain manifold we assume spherical symmetry and use
Schwarzschild coordinates
\begin{equation}
g_{ab}dx^a dx^b= -e^{-2 \delta} A\: dt^2 + A^{-1} dr^2 + r^2
(d\theta^2+\sin^2{\!\theta}\: d\phi^2),
\end{equation}
where $\delta$ and $A$ are functions of $(t,r)$.
Next, we assume that the wave maps are corotational, that is
\begin{equation}
F= F(t,r),\quad \Theta=\theta, \quad \Phi=\phi.
\end{equation}
Equation (3) reduces then to the single semilinear wave equation
\begin{equation}
\square_g F - \frac{\sin(2 F)}{r^2} = 0,
\end{equation}
where
\begin{equation}
\square_g = -e^{\delta} \partial_t(e^{\delta} A^{-1} \partial_t)
+\frac{e^{\delta}}{r^2} \partial_r(r^2 e^{-\delta} A\:
\partial_r),
\end{equation}
and the Einstein equations become
\begin{eqnarray}
{\partial_t A} &=& -2 \alpha\: r A (\partial_t F) (\partial_r F),
\\
{\partial_r \delta} &=& -\alpha\: r \left((\partial_r F)^2 +
A^{-2} e^{2\delta} (\partial_t F)^2 \right),
\\
\partial_r A &=& \frac{1-A}{r} - \alpha\: r \left( A (\partial_r F)^2 + A^{-1}
e^{2\delta} (\partial_t F)^2 + 2 \:\frac{\sin^2{\!F}}{r^2}\right).
\end{eqnarray}
These equations are invariant under dilations $(t,r) \rightarrow
(\lambda t, \lambda r)$ so it is natural to look for continuously
self-similar (CSS) solutions, that is solutions which are left
invariant by the action of the homothetic Killing vector $K=
t\partial_t+r\partial_r$. To study such solutions it is convenient
to use   similarity variables $\rho=r/(-t)$ and $\tau=-\ln(-t)$.
Then $K=-\partial_{\tau}$, so CSS solutions do not depend on
$\tau$. Assuming this and using an auxiliary function
$Z=e^{\delta} \rho/A$, we reduce equations (8-12)  to the system of
ordinary differential equations (where prime is $d/d\rho$)
\begin{eqnarray}
F''&+& \frac{2}{\rho} F' - \alpha(1+Z^2)\rho {F'}^3  - \frac{\sin(2
F)}{A \rho^2(1-Z^2)} = 0,\\
A'&=& - 2 \alpha \rho A {F'}^2,\\
\rho Z' & =& Z (1+\alpha (1-Z^2) \rho^2 {F'}^2),\\
\rho A' &=& 1-A -\alpha \left(\rho^2 A (1+Z^2) {F'}^2 + 2
\sin^2{F}\right).
\end{eqnarray}
The combination of (14) and (16)  yields the constraint
\begin{equation}
1-A  - 2 \alpha \sin^2{F} +\alpha A \rho^2 {F'}^2 (1-Z^2) =0.
\end{equation}
This system of equations has a fixed singularity at the center
$\rho=0$ and moving singularities at points where $Z(\rho)=\pm 1$
and/or $A(\rho)=0$. In terms of the similarity coordinate $\rho$,
the metric (6) takes the  form
\begin{equation}
 ds^2 = A^{-1} (1-Z^{-2})\:\rho^2  dt^2 + 2 A^{-1} t \rho\: dt d\rho
 + A^{-1} t^2 d\rho^2 + t^2 \rho^2 (d\theta^2+\sin^2{\!\theta}\: d\phi^2),
\end{equation}
hence the hypersurfaces $Z=\pm 1$ are null (provided that $A>0$).
The first $\rho_1$ where $Z(\rho_1)=1$ is the locus of the past
light cone of the singularity at the origin $(t=0,r=0)$ (in what
follows we shall refer to the past and future light cones of the
singularity as to the past and future self-similarity horizons
(SSH)). By rescaling, $\rho \rightarrow \rho/\rho_1$, one can
always locate the past self-similarity horizon at $\rho_1=1$, that
is $Z(1)=1$. To ensure regularity of solutions in the interval
$0\leq\rho\leq 1$, the equations (13-17) must be supplemented by
the boundary conditions at both endpoints
\begin{eqnarray}
F(0)&=&0,\,\quad F'(0)=c, \quad Z(0)=0, \quad A(0)=1, \\ \label{ics0}
F(1)&=&\frac{\pi}{2},\quad F'(1)=b, \quad Z(1)=1, \quad A(1)=1-2\alpha,
\end{eqnarray}
where $c$ and $b$ are free parameters.

Our main result is the following theorem:
\begin{thm} For any $0 \leq \alpha <1/2$ and any nonegative
integer $n$, the equations (13-17)  have an analytic solution
$(F_n, A_n, Z_n)$ which satisfies the boundary conditions (19-20)
and has precisely $n$ oscillations of $F_n(\rho)$ around $\pi/2$.
\end{thm}
In the next section we shall prove this theorem using a shooting
technique. The numerical evidence for Theorem~1 was given in I  .
The case $\alpha=0$ was proved previously in~\cite{bi} so
hereafter we assume that $0 < \alpha <1/2$.
\section{Proof of Theorem~1}
For convenience we rewrite equations (13-15)
in terms of $H=F-\pi/2$:
\begin{eqnarray}
H''&+& \frac{2}{\rho} H' - \alpha(1+Z^2)\rho {H'}^3  + \frac{\sin(2
H)}{A \rho^2(1-Z^2)} = 0, \label{eqH}\\
A'&=& - 2 \alpha \rho A {H'}^2,\label{eqA}\\
\rho Z' & =& Z (1+\alpha (1-Z^2) \rho^2 {H'}^2) \label{eqZ},
\end{eqnarray}
The constraint becomes
\begin{equation}\label{const}
1-2\alpha - A  + 2 \alpha \sin^2\!{H} +\alpha A \rho^2 {H'}^2 (1-Z^2) =0.
\end{equation}
The initial conditions at $\rho=0$ are
\begin{equation}\label{ics}
H(0)=-\frac{\pi}{2}, \quad H'(0)=c, \quad A(0)=1, \quad Z(0)=0, \quad Z'(0)=1.
\end{equation}
Note that the above equations have a residual scaling symmetry
$\rho \rightarrow \lambda \rho$. The initial condition $Z'(0)=1$
is imposed temporarily in order to fix the scale. We shall refer
to solutions of equations (21-24) satisfying  the initial
conditions (\ref{ics}) as to  $c$-orbits. In the appendix we show
that $c$-orbits exist locally and are analytic in $\rho$ and $c$.
Now we shall show that $c$-orbits can be extended up to a point
$\rho_1$ at which $Z(\rho_1)=1$.
\begin{prop}
For any $0 < \alpha <1/2$ and $c>0$ there is a $\rho_1(c) \in (\sqrt{1-2\alpha},1)$,
such that the $c$-orbit
is defined for all $\rho<\rho_1$ and  $\lim_{\rho \rightarrow \rho_1} Z(\rho)=1$.
Furthermore, the following limits exist
$$
-\frac{\pi}{2}< \bar H \stackrel{def}{=} \lim_{\rho \rightarrow
\rho_1} H(\rho)<\frac{\pi}{2}, \quad \bar A \stackrel{def}{=}
\lim_{\rho \rightarrow \rho_1} A(\rho) =1-2\alpha \cos^2{\bar H},
\quad \lim_{\rho \rightarrow \rho_1} (1-Z^2) {H'}^2=0.
$$
\end{prop}
\noindent \emph{Proof:} Let the maximum domain of definition of
the $c$-orbit be $0\leq \rho <\rho_1$ and assume that $Z(\rho)<1$
in this interval. Then, from constraint (\ref{const}) we have $A
\geq 1-2\alpha>0$ and hence $\bar A= \lim_{\rho \rightarrow
\rho_1} A(\rho) > 0$ ($\bar A$ exists since $A(\rho)$ is monotone
decreasing). By (\ref{eqZ}) $Z' \geq 0$, hence $\bar Z =\lim_{\rho
\rightarrow \rho_1} Z(\rho)$ exists.  If $\bar Z<1$, then from
constraint (\ref{const}) ${H'}^2$ is bounded so $\bar H=\lim_{\rho
\rightarrow \rho_1} H(\rho)$ exists, which in turn implies, again
by (\ref{const}), that $\lim_{\rho \rightarrow \rho_1} H'$ exits.
Thus, $H, H', A$, and $Z$ all have finite limits at $\rho_1$ and
therefore the $c$-orbit may be continued beyond $\rho_1$
contradicting the maximality of $\rho_1$. We conclude that $\bar
Z=1$.

Now, we must show that $\bar H \in (-\pi/2,\pi/2)$ exists. Since
$\bar Z =1$, we may no longer conclude that ${H'}^2$ is bounded
but from equation (\ref{eqA}) we get $(\ln{A})'=-2\alpha \rho
{H'}^2$, so ${H'}^2$ is integrable near $\rho_1$ which implies
that  $H'$ is absolutely integrable ($|H'|<1+{H'}^2$) and thus
$\bar H$ exists. From constraint (\ref{const}), $H(\rho)=\pm
\pi/2$ for some $0<\rho<\rho_1$ is not possible since $1-A>0$.
Thus, $-\pi/2<H(\rho)<\pi/2$ and so $-\pi/2\leq \bar H \leq
\pi/2$. In fact, for $\rho \geq \rho_1/2$ we have $1-A \geq
\sigma>0$, so $2\alpha \cos^2{H}\geq \sigma>0$ (remember that we
assume $\alpha>0$), hence $H$ is uniformly bounded away from $\pm
\pi/2$, and thus $-\pi/2< \bar H < \pi/2$.

To prove $\bar A=1-2\alpha \cos^2{\bar H}$, note that by
(\ref{const}) $d=\lim_{\rho \rightarrow \rho_1} {H'}^2 (1-Z^2)$
exits and  is finite. Hence, by (\ref{eqZ}) $\lim_{\rho
\rightarrow \rho_1} Z'$ exists and is finite so
$1-Z^2=O(\rho-\rho_1)$ near $\rho_1$. If $d\neq 0$, then $H'(\rho)
\sim d/(\rho_1-\rho)$ would not be integrable near $\rho_1$, thus
$d$  must be zero. Inserting this into (\ref{const}) we get $\bar
A =1-2\alpha \cos^2\!{\bar H}$.

Next, $(Z/\rho)'>0$ by (\ref{eqZ}) and $\lim_{\rho\rightarrow 0} (Z/\rho)=1$ by
L'H\^opital's rule, hence $Z\geq \rho$ for all $\rho>0$, and thus $\rho_1\leq
1$. Finally, from (\ref{eqA}) and (\ref{eqZ})
\begin{equation}
\left( \frac{A Z^2}{\rho^2}\right)' =-\frac{2 Z^4 A \alpha {H'}^2}{\rho} <0,
\end{equation}
and since $\lim_{\rho\rightarrow 0} (A Z^2/\rho^2)=1$, we have $(A
Z^2/\rho^2)\leq 1$ and hence $\rho_1>\sqrt{A}>\sqrt{1-2\alpha}$.

If $Z(\rho_2)=1$ for some $\rho_2<\rho_1$, we replace $\rho_1$ by $\rho_2$ in
the above arguments.

\begin{cor}
The function $\rho_1(c)$ is continuous.
\end{cor}
\noindent \emph{Proof:} Let $\tilde c$ be given and let
$\epsilon>0$. By Proposition~2, $\rho_1(\tilde c)$ is defined. The
function $Z(\rho)$ is monotone increasing for $\rho<\rho_1(\tilde
c)$, so $Z(\rho_1(\tilde c)-\epsilon,\tilde c)<1$, hence for all
$c$ sufficiently close to $\tilde c$, $Z(\rho_1(\tilde
c)-\epsilon,c)<1$, and thus $\rho_1(c)>\rho_1(\tilde c)-\epsilon$.
To show that $\rho_1(c)<\rho_1(\tilde c)+\epsilon$ for all $c$
sufficiently close to $\tilde c$, we assume otherwise and get a
contradiction. By the mean-value theorem $Z(\rho_1(\tilde
c)+\epsilon,c)-Z(\rho,c)=Z'(\xi,c) (\rho_1(\tilde
c)+\epsilon-\rho)$. By continuity $Z(\rho,c)$ is close to
$Z(\rho,\tilde c)$  and $Z(\rho,\tilde c)$ is close to $1$ if
$\rho$ is close to $\rho_1(\tilde c)$, hence $Z(\rho,c)$ is
arbitrarily close to $1$. But $Z'(\rho,c)>Z(\rho,c)/\rho>1$, so
$Z(\rho_1(\tilde c)+\epsilon,c)>Z(\rho,c)+\epsilon>1$, which is a
contradiction. Thus, $\rho_1(c)<\rho_1(\tilde c)+\epsilon$.
\vskip 0.1cm
\begin{lem}
$H'(\rho)$ is bounded near $\rho_1$ if and only if $\bar H=0$.
\end{lem}
\noindent \emph{Proof:} Suppose that $\bar H \neq 0$ and
$H'(\rho)$ is bounded. Then, in (21) we have
\begin{equation}
H''=\mbox{bounded terms} - \frac{\sin{2 H}}{A \rho^2 (1-Z^2)} \sim
\frac{d}{\rho_1-\rho},
\end{equation}
where $d \neq 0$. This contradicts that  $H'(\rho)$ is bounded
near $\rho_1$ and concludes  the "only if" part of Lemma~4.

 Suppose now that $
H(\rho_1)=0$ and $H'(\rho)$ is unbounded. Without loss of
generality we consider the case that $H(\rho)<0$ and $H'(\rho)>0$
near $\rho_1$. Dividing equation (\ref{eqH}) by $H'$ and
integrating from $\rho$ to $\rho_1$ we obtain
\begin{equation}\label{iden0}
\int_{\rho}^{\rho_1} \left( \frac{H''}{H'} + \frac{2}{\rho}  -
\alpha(1+Z^2)\rho {H'}^2 + \frac{\sin(2 H)}{H' A
\rho^2(1-Z^2)}\right) d\rho= 0.
\end{equation}
The first integral is divergent because $\lim_{\rho \rightarrow
\rho_1} \ln{H'} =\infty$. The second and the third terms are
integrable (remember that ${H'}^2$ is integrable). Thus, to get a
contradiction it suffices to show that the last term is
integrable. We write this term as
\begin{equation}\label{ss1}
\frac{\sin(2 H)}{H' A \rho^2(1-Z^2)} = \frac{\sin(2 H)} {H A
\rho^2} \frac{H}{(1-Z^2) H'}.
\end{equation}
The first factor is continuous and we now show that the second
factor is also continuous. Applying L'H\^opital's rule we get
\begin{equation}\label{lop}
\lim_{\rho\rightarrow\rho_1} \frac{H}{(1-Z^2) H'} =
\lim_{\rho\rightarrow\rho_1}  \frac{H'}{-2 Z Z' H' +(1-Z^2) H''} =
\lim_{\rho\rightarrow\rho_1}  \frac{1}{-2 Z Z'  +(1-Z^2) H''/H'}.
\end{equation}
Next, using (\ref{eqH}) we get
\begin{equation}\label{lop2}
(1-Z^2)\frac{H''}{H'} = -\frac{2 (1-Z^2)}{\rho} +\alpha \rho
(1+Z^2)(1-Z^2) {H'}^2 -\frac{\sin(2 H)}{A \rho^2 H'}.
\end{equation}
In the limit $\rho\rightarrow\rho_1$, the  first term on the rhs
of (\ref{lop2}) obviously goes to zero, the second does by
Proposition~2, and the third does by the assumption that
$H'\rightarrow \infty$. Thus, the limit (\ref{lop}) is finite and
consequently so is (\ref{ss1}). This contradicts (\ref{iden0}) and
thus concludes the proof of the "if" part of Lemma~4.
\begin{cor} A $c$-orbit which has $\bar H(c)$=0 is analytic on
the whole interval $0\leq \rho \leq \rho_1$.
\end{cor}
\noindent{\emph{Proof:}}
 The boundedness of $H'(\rho)$ implies by (\ref{eqH}) that
$H''>-2 H'/\rho$ is bounded below (remember that  $H(\rho)<0$ and
$H'(\rho)>0$ near $\rho_1$), hence $\lim_{\rho\rightarrow \rho_1}
H'(\rho)$ exists. Having that, it is easy to show by applying
L'H\^opital's rule  to $\lim_{\rho\rightarrow \rho_1}
(H^{(k)}(\rho_1)-H^{(k)}(\rho))/(\rho_1-\rho)$ for $k=0,1$ that
the solution $(H,A,Z)$ is $C^2$ near $\rho_1$. By a routine
contraction mapping argument one can show that $C^2$ solutions are
unique, hence a $c$-orbit must belong to the one-parameter family
of analytic solutions from Proposition~14 (see the appendix).
\vskip 0.15cm
 Next, we describe the behaviour of
$c$-orbits for small and large values of the shooting parameter
$c$. We define  a nodal number of a $c$-orbit $N(c)=$ number of
zeros of the function $H(\rho)$ on the interval $0 \leq \rho <
\rho_1$. We first show that $c$-orbits with small $c$ have no
nodes.
\begin{prop}
If $c$ is sufficiently small then $N(c)=0$.
\end{prop}
\noindent \emph{Proof:} For $c=0$ we have $H(\rho)\equiv -\pi/2$
and $Z(\rho)=\rho$ so $\rho_1(c=0)=1$. By continuity, for any
$\epsilon>0$ and sufficiently small $c$ we can find $\rho_0$ such
that $1-\epsilon<\rho_0<\rho_1(c)<1$ and
$H(\rho_0)<-\pi/2+\epsilon$. We know from the proof of
Proposition~2 that $\lim_{\rho\rightarrow \rho_1}
\sqrt{\rho_1-\rho}\: H'=0$, hence
\begin{equation}\label{proof6}
H(\rho_1)-H(\rho_0)=\int_{\rho_0}^{\rho_1} H'(\rho) d\rho < const
\int_{\rho_0}^{\rho_1} \frac{d\rho}{\sqrt{\rho_1-\rho}}  < const
\sqrt{\epsilon}.
\end{equation}
Thus,  $H(\rho)$ stays  arbitrarily close to $-\pi/2$  all the way
up to $\rho_1$ if $c$ is  sufficiently small and therefore
$N(c)=0$. We remark that using a scaling argument one can derive
the precise asymptotic behaviour of $c$-orbits for small $c$. We
omit this argument since it is not needed for the proof. \vskip
0.15cm
We show next that $c$-orbits with large $c$ have arbitrarily many nodes.
\begin{prop}
$N(c) \rightarrow \infty$ for $c \rightarrow \infty$.
\end{prop}
\noindent \emph{Proof:}  We rescale the variables, setting
\begin{equation}
x=c \rho, \quad \tilde H(x)=H(\rho), \quad \tilde A(x)=A(\rho), \quad
\tilde Z(x) = c Z(\rho).
\end{equation}
Then, equations (21-24) become
\begin{eqnarray}
\tilde H''&+ &\frac{2}{x} \tilde H' -
\alpha(1+\frac{\tilde Z^2}{c^2} )x {H'}^3  + \frac{\sin(2
\tilde H)}{\tilde A x^2 (1-\frac{\tilde Z^2}{c^2})} = 0, \\
\tilde A'&=& - 2 \alpha x \tilde A {\tilde{H'}}^2,\\
x \tilde Z' & =& \tilde Z (1+\alpha (1-\frac{\tilde Z^2}{c^2}) x^2
{\tilde{H'}}^2),
\end{eqnarray}
with  the constraint
\begin{equation}
1-2\alpha - \tilde A  + 2 \alpha \sin^2\!{\tilde H} +\alpha \tilde
A x^2 {\tilde{H'}}^2 (1-\frac{\tilde{Z^2}}{c^2}) =0,
\end{equation}
and  the initial conditions at $x=0$
\begin{equation}
\tilde H(0)=-\frac{\pi}{2}, \quad \tilde H'(0)=1, \quad \tilde A(0)=1,
\quad \tilde Z(0)=0, \quad \tilde Z'(0)=1.
\end{equation}
As $c \rightarrow \infty$, the solutions of equations (34)-(38)
tend uniformly on compact intervals to solutions of the limiting
equations
\begin{eqnarray}
h''&+ &\frac{2}{x} h' -
\alpha x {h'}^3  + \frac{\sin(2 h)}{a x^2} = 0, \label{eqh}\\
a'&=& - 2 \alpha x a {h'}^2,\label{eqa} \\
x z' & =& z (1+\alpha
x^2 {h'}^2),\label{eqz}
\end{eqnarray}
with  the constraint
\begin{equation}
1-2\alpha - a + 2 \alpha \sin^2\!{h} +\alpha a
x^2 {h'}^2 =0, \label{cons}
\end{equation}
and the same  initial conditions at $x=0$
\begin{equation}\label{ics_lim}
h(0)=-\frac{\pi}{2}, \quad h'(0)=1, \quad a(0)=1, \quad z(0)=0, \quad z'(0)=1.
\end{equation}
We observe first that the function $a(x)$ is monotone decreasing
by (\ref{eqa}) and  bounded below, $a>1-2\alpha$, by (\ref{cons}).
Thus, no singularity can develop due to $a$ going to zero. Also,
by (\ref{cons}) no singularity can develop due to $h'$ becoming
unbounded. Thus, solutions exist for all $x>0$ (assuming the
existence of a solution for small $x$). In order to complete the
proof it is sufficient to show that the function $h(x)$ has an
infinite number of zeros for $x>0$. Since $a<1$, it follows from
(\ref{cons}) that $-\pi/2<h(x)<\pi/2$ for all $x>0$.
 To show that $h(x)$ oscillates around zero
we consider three cases:\\
(i) Assume that $\lim_{x \rightarrow \infty} h(x)$ does not exist. Then, there must be a
sequence
$\ldots x_k<y_k<x_{k+1}<y_{k+1}<\ldots$ such that $h$ has a local minimum at $x_k$ and a local
maximum at $y_k$. By (\ref{eqh}), $h'(x_k)=0, h''(x_k) \geq0$ imply that $\sin(2 h(x_k))\leq
0$, hence $h(x_k)\leq 0$. By a similar argument, $h(y_k) \geq 0$. Thus, $h(x)$
has a zero in each interval $x_k<x<y_k$. \\
(ii) Assume that a nonzero $\lim_{x \rightarrow \infty} h(x)$
exists. Then, from (\ref{cons}) $\lim_{x \rightarrow \infty} x^2
{h'}^2$ exists and, in fact, equals zero because
 $\lim_{x \rightarrow \infty} h(x)$ exists. This implies by (\ref{eqh}) that
 $\lim_{x \rightarrow \infty} x^2 h''(x) = -\sin(2 h(\infty))/A(\infty) \neq 0$, hence
 $\lim_{x \rightarrow \infty} x^2 {h'}^2(x) \neq 0$. Thus the case (ii) does no arise.\\
\noindent (iii) Assume that $\lim_{x \rightarrow \infty} h(x)= 0$. We define the rotation function
$\theta(x)$ by
\begin{equation} \label{rot}
\tan{\theta(x)} = \frac{x h'(x)}{h(x)}, \qquad  \theta(0) = 0.
\end{equation}
\emph{Remark 1.} The rotation function $\theta(x)$ is well defined because the case
$h(x)=h'(x)=0$ is impossible for solutions satisfying the initial conditions (\ref{ics_lim}). To see
this, assume that $h(x_0)=h'(x_0)=0$ for some $x_0>0$. Then, by (\ref{cons}) $a(x_0)=1-2\alpha$
and the unique solution with these initial conditions at $x_0$  is $h(x)=0, h'(x)=0,
a(x)=1-2\alpha$ for all $x$, contradicting the initial conditions
(\ref{ics_lim}).
\vskip 0.2cm
\noindent  We want to show that $\lim_{x \rightarrow \infty} \theta(x)=-\infty$.
Using (\ref{eqh}) we obtain
\begin{equation}
x \theta'(x)= -\sin^2{\theta} - \frac{\sin{2 h}}{2 h} \frac{2\cos^2\!{\theta}}{a}
- \frac{(1-2\alpha \cos^2{h})\sin{\theta}\cos{\theta}}{a}.
\end{equation}
Under the assumption $\lim_{x \rightarrow \infty} h(x) = 0$, it follows from (\ref{cons}) that
$\lim_{x \rightarrow \infty} a(x) =1-2\alpha$, hence for sufficiently large $x$
\begin{equation}
\theta'(x) \approx  -\frac{1}{x} \left( \sin^2{\theta} + \sin{\theta} \cos{\theta} +
\frac{2\cos^2{\theta}}{1-2\alpha}\right) < -\frac{3}{4 x},
\end{equation}
so $\lim_{x \rightarrow \infty} \theta(x)=-\infty$. Thus, given
any integer $k$ there exists an $x_k$ such that $h(x)$ has at
least $k$ zeroes for $x<x_k$. By continuous dependence on initial
conditions, we may choose $c>x_k/\sqrt{1-2\alpha}$ so that the
$c$-solution has $k$ zeroes also for $x<x_k$. In terms of the
variable $\rho=x/c$ the $c$-solution has $k$ zeroes for
$\rho<\sqrt{1-2\alpha}<\rho_1(c)$. This completes the proof of
Proposition~7. \vskip 0.15cm Next, we need two lemmas which tell
us how the number of nodes $N(c)$ changes under small variations
 of $c$.
\begin{lem}
If $\bar H(\tilde c) \!= \!0$, then $N(c)\!=\!N(\tilde c)$ or $N(c)\!=\!N(\tilde c)+1 $ for $c$ sufficiently
close to $\tilde c$.
\end{lem}
\noindent \emph{Proof:} First note that if $H(\rho,\tilde c)$ has
a zero at some $\rho_0<\rho_1(\tilde c)$, then $H'(\rho_0,\tilde
c) \neq 0$ (see Remark~1) so by continuity of $H(\rho,c)$ with
respect to $c$, $H(\rho,c)$ also has a zero if $c$ is sufficiently
close to $\tilde c$. Thus $N(c)\geq N(\tilde c)$ and it suffices
to show that $N(c)\leq N(\tilde c)+1$. Let $\tilde a<\rho_1(\tilde
c)$ be the last node of the $\tilde c$-orbit, that is $H(\tilde
a,\tilde c)=0$ and, for concreteness, $H(\rho,\tilde c)<0$ for
$\tilde a<\rho<\rho_1$. By continuity with respect to $c$,
 $H(\rho,c)$ will also have a zero at $a$ near $\tilde a$ if  $c$ is near
$\tilde c$.  In order to prove that $H(\rho,c)$
cannot have more than one zero in the interval $a<\rho<\rho_1(c)$, we now show  that if $H(\rho,c)$
becomes positive for some $\rho>a$, then it would not have time to change the sign again before going singular.
Assume for contradiction that there is
 a segment $a<\rho_N\leq \rho\leq\rho_D$ of the $c$-orbit
in which the function
$H(\rho)$ is monotone decreasing from a local maximum $H(\rho_N)>0$ to $H(\rho_D)=0$.

\noindent We define
\begin{equation}
W=\frac{1}{2} \rho^2 A {H'}^2 (1-Z^2)+\sin^2{H
}.
\end{equation}
From (\ref{const})
$W=(A-1+2\alpha)/(2 \alpha)$, hence by (\ref{eqA}) $W'<0$.
We have
\begin{equation}\label{w1}
\frac{{H'}^2}{W-\sin^2{H}} =\frac{2}{\rho^2 A (1-Z^2)}, \quad \mbox{so} \quad
\frac{-H'}{\sqrt{W-\sin^2\!{H}}} =\frac{\sqrt{2}}{\rho \sqrt{A (1-Z^2)}}.
\end{equation}
Integrating the left-hand side from $\rho_N$ to $\rho_D$, we get (using $H_N=H(\rho_N)$)
\begin{equation}\label{w2}
\int_{\rho_N}^{\rho_D} \frac{-H' d\rho }{\sqrt{W-\sin^2{H}}}=
\int_{0}^{H_N} \frac{dH}{\sqrt{W-\sin^2\!{H}}}\geq
\int_{0}^{H_N} \frac{dH}{\sqrt{\sin^2\!{H_N}-\sin^2\!{H}}} > \frac{\pi}{2},
\end{equation}
where the first inequality follows from $W(\rho)\leq W(\rho_N)=\sin^2\!{H_N}$
(since $W'$ decreases) and the second inequality is a simple calculation using a substitution $\sin{H}=u
\sin{H_N}$ (remember that $H_N<\pi/2$).

Next, we derive an upper bound for the integral of the right-hand side of (\ref{w1}). We have
\begin{equation}
\int_{\rho_N}^{\rho_D} \frac{d\rho}{\rho \sqrt{A (1-Z^2)}} \leq \frac{1}{\rho_N\sqrt{1-2\alpha}}
\int_{\rho_N}^{\rho_D} \frac{d\rho}{\sqrt{1-Z^2}} \leq \frac{1}{\rho_N\sqrt{1-2\alpha}}
\int_{\rho_N}^{\rho_D} \frac{d\rho}{\sqrt{1-Z}}.
\end{equation}
We showed above that $Z'>1$, hence $1-Z \geq \rho_1-\rho$. Therefore
\begin{equation}
\int_{\rho_N}^{\rho_D} \frac{d\rho}{\sqrt{1-Z}} \leq
\int_{\rho_N}^{\rho_D} \frac{d\rho}{\sqrt{\rho_1-\rho}} = 2 (\sqrt{\rho_1-\rho_N}-\sqrt{\rho_1-\rho_D})
<2 \sqrt{\rho_1-\rho_N}.
\end{equation}
By continuity of solutions with respect to $c$ and
 by Corollary~3,
$\rho_N$ is arbitrarily close to $\rho_1(c)$ if $c$ is
sufficiently close to $\tilde c$, hence it follows from (51) that
 the integral of
the right-hand side of (\ref{w1}) is arbitrarily small. This is in
contradiction with (\ref{w2}), hence $H(\rho,c)$ cannot have a
second additional zero, which completes the proof of Lemma~8.
\begin{lem}
If $\bar H(\tilde c) \neq 0$, then $N(c)=N(\tilde c)$ for $c$ sufficiently close to $\tilde c$.
\end{lem}
\noindent \emph{Proof:}
Without loss of generality we assume that
$\bar H(\tilde c)<0$. As above
let $\tilde a<\rho_1(\tilde c)$ be the last node of the $\tilde c$-orbit, that is $H(\tilde a,\tilde c)=0$
and $H(\rho,\tilde c)<0$ for $\tilde a<\rho\leq \rho_1$. Let $a$ be the corresponding zero of $H(\rho,c)$ for $c$
near $\tilde c$. We want to show that $H(\rho,c)$ cannot have an extra zero for $\rho>a$.
 Suppose for contradiction that $H(b,c)=0$ for some $b>a$. Then, there must be a $\delta<b$ such that
 $H(\delta,c)=\bar H(\tilde c)$.
Let us integrate the identity
\begin{equation}\label{iden}
\frac{H'}{\sqrt{W-\sin^2\!{H}}} =\frac{\sqrt{2}}{\rho \sqrt{A (1-Z^2)}}
\end{equation}
from $\delta$ to $b$. For the left hand side we get
\begin{equation}
\int_\delta^b \frac{H' d\rho }{\sqrt{W-\sin^2{H}}}=
\int_{0}^{-\bar H} \frac{dH}{\sqrt{W-\sin^2\!{H}}}.
\end{equation}
From Proposition~2 we know that $\lim_{\rho \rightarrow \rho_1}
(1-Z^2) {H'}^2=0$ so $W(\rho,\tilde c)<(1+\epsilon/2) \sin^2{\bar
H}$ for $\rho$ near $\rho_1$ and hence $W(\rho, c)<(1+\epsilon)
\sin^2{\bar H}$ for $c$ near $\tilde c$. Since $W$ is decreasing,
$W(\delta, c)< W(\rho,c) < (1+\epsilon) \sin^2{\bar H}$. Thus
\begin{equation}
\int_{0}^{-\bar H} \frac{dH}{\sqrt{W-\sin^2\!{H}}} \geq
\int_{0}^{-\bar H} \frac{dH}{\sqrt{(1+\epsilon) \sin^2\!{\bar H} -\sin^2\!{H}}}
\geq \arcsin\left(\frac{1}{\sqrt{1+\epsilon}}\right) > \frac{\pi}{2}
\end{equation}
for sufficiently small $\epsilon$, where the last but one inequality can be seen by substituting
$\sin{H}= u \sin{\bar H}$ into the integral. By the same argument as in  the proof of
 Lemma~8, the integral of the right
hand side of (\ref{iden}) is $O(\sqrt{\rho_1-\rho})$. By
continuity of solutions with respect to $c$ and
 by Corollary~3, $\delta$ is arbitrarily close to $\rho_1(c)$ if
$c$ is sufficiently close to $\tilde c$, hence the integral of the
left hand side of equation (52) is arbitrarily small. This
contradicts (54) and completes
 the proof of Lemma~9.
\vskip 0.15cm
Now we are ready to make a shooting argument. We define a set
\begin{equation}
C_0=\{ c \:|\:  N(c) =0 \}
\end{equation}
and let  $c_0=\sup C_0$. The set $C_0$ is nonempty (by
Proposition~6) and bounded above (by Proposition~7)
 so $c_0$ exists.
We claim that the $c_0$-orbit has no nodes and satisfies the boundary condition $\bar H(c_0)=0$.
To see this, note that the $c_0$-orbit cannot have a node because then by
 Lemmas~8 and 9 all nearby $c$-orbits would have
 a node
so there would be an interval around $c_0$ without any elements of $C_0$ in it,
 contradicting
the assumption that $c_0$ is the least upper bound. Thus,
$N(c_0)=0$. Now, if $\bar H(c_0)<0$, then by Lemma~9 all nearby
$c$-orbits would have no nodes, so there would be an interval
around $c_0$ consisting of elements of $C_0$, contradicting the
assumption that $c_0$ is an upper bound of $C_0$. Thus $\bar
H(c_0)=0$.

 Next, we define $C_1=\{ c>c_0\: |\:  N(c) =1 \}$.
This set is nonempty by the previous step and Lemma~8 and bounded
above by Proposition~7, hence $c_1=\sup C_1$ exists. By the same
argument as above, the $c_1$-orbit has  exactly one node and
satisfies $\bar H(c_1)=0$. The construction of subsequent
$c_n$-orbits proceeds by induction. \vskip 0.1cm \noindent
\emph{Conclusion of  the proof of Theorem~1:}

Returning to the original variable $F(\rho)$ and rescaling
$\rho\rightarrow \rho/\rho_1(c_n)$ we get the solution of
equations (13-17) which satisfies the boundary  conditions (19)
and (20) and has exactly $n$ intersections with the line
$F=\pi/2$. By Corollary~5 this solution is analytic in the whole
interval $0\leq \rho\leq 1$.
\section{Beyond the past self-similarity horizon}
In this section we consider the behaviour of the CSS solutions of
Theorem~1 outside the past SSH, in particular we ask the question:
do these solutions possess a regular future self-similarity
horizon? Note that $\rho=\infty$ corresponds to the hypersurface
$(t=0, r>0)$ so in order to analyze the global behaviour of
solutions (for $t>0$) we need to go "beyond $\rho=\infty$". To
this end we define, after I, a new coordinate $x$ by
\begin{equation}
\frac{d}{dx}=\rho Z \:\frac{d}{d \rho}, \qquad x(\rho=1)=0.
\end{equation}
We also define an auxiliary function $w(x)=1/Z(\rho)$. In these
new variables, the past SSH where $w=1$ is at $x=0$, while the
future SSH (if it exists) is  at some $x_A>0$ where $w(x_A)=-1$.

In terms of $x$ and $w$, the equations (21)-(23) become autonomous
(where now prime is $d/dx$)
\begin{eqnarray}
H''& -& 2\alpha w {H'}^3 +  \frac{\sin(2 H)}{A(w^2-1)} = 0,\\
A'&=& -2 \alpha A w {H'}^2, \\
w' &=& -1 +\alpha (1-w^2) {H'}^2.
\end{eqnarray}
The constraint (24) becomes
\begin{equation}
1-2\alpha - A  + 2 \alpha \sin^2{H} +\alpha A {H'}^2 (w^2-1) =0.
\end{equation}
 From (20) the initial conditions at $x=0$ are
\begin{equation}\label{ics3}
H(x) \sim  b x,\quad w(x) \sim 1 -x, \quad A(x) \sim 1-2\alpha - 2
\alpha (1-2 \alpha) b^2 x.
\end{equation}
We  know from Theorem~1 that for each $\alpha<1/2$ there is an
infinite sequence $\{b_n(\alpha)\}$ determining solutions which
are regular inside the past SSH, that is for all $x\leq 0$ (note
that $\rho=0$ corresponds to $x=-\infty$).  In I we showed that
for $x>0$ the solutions starting from the past SSH with the
initial conditions (\ref{ics3}) tend in finite "time" to $w=-1$ if
$b$ is small, or to $w=+1$ if $b$ is large. After I we shall refer
to these two kinds of behaviour as to type A and type B solutions,
respectively.
 Now we want to show that the solutions of Theorem~1 are of type A
(and therefore possess the future SSH)
 provided that $\alpha$ is sufficiently small.
Unfortunately, the shooting argument gives us insufficient
information about the parameters $b_n$ so we cannot apply the
above mentioned result of I to determine the character of
solutions outside the past SSH. Instead, we shall make use of the
obvious fact that for $\alpha=0$ all solutions are of type A.
\begin{lem} For sufficiently small
$\alpha$ the $c_n$-orbits of Theorem~1 (rescaled so that
$\rho_1(c)=1$) have $|b_n|$ uniformly bounded above for all $n$.
\end{lem}
\noindent \emph{Proof:} It was shown in~\cite{bi} (see Lemma~4 in
that reference) that for $\alpha=0$ the solution to equations
(57)-(61) for $x<0$ must exit the strip $|H|\leq \pi/2$  if $|b|$
is too large, say $|b|>B$. By continuous dependence, the same is
true for sufficiently small $\alpha$. But from Proposition~2 the
$c$-orbits must stay in the strip $|H|\leq \pi/2$ for all $x<0$.
Thus, $|b_n|\leq B$ for small $\alpha$.
\begin{lem}
If a solution to equations (57)-(60) has $w(x_0)<0$ and
$A(x_0)>2/3$ for some $x_0$, then there is $x_A>x_0$ such that
$\lim_{x\rightarrow x_A} w(x)=-1$, i.e., the solution is of type
A.
\end{lem}
\noindent \emph{Proof:} By (58) $A$ is increasing for $w<0$. Thus,
 using equation (59) and
the constraint (60) we get for $x>x_0$
\begin{equation}
  w'=-1+\alpha (1-w^2) {H'}^2 = -1 + \frac{1-A-2\alpha \cos^2{H}}{A}
  < -2+\frac{1}{A} \leq -\frac{1}{2},
\end{equation}
hence $w$ must hit $-1$ for some finite $x_A>x_0$.
\begin{prop}
The $c_n(\alpha)$-orbits are of type A if $\alpha$ is sufficiently
small.
\end{prop}
\noindent \emph{Proof:} For $\alpha=0$ and any $b$ we have $
w(x)=1-x$ and $A(x)\equiv 1$; in particular $A(3/2)=1>2/3$ and
$w(3/2)=-1/2<0$. By continuous dependence on initial conditions
there exists a $\delta(b)$ such that if $\alpha<\delta(b)$ and
$|b-b'|<\delta(b)$, then $A(3/2,b')>2/3$ and $w(3/2,b')<0$. This
implies by Lemma~11 that the solutions corresponding to such
values of $\alpha$ and $b'$ are of type A. By a standard theorem
of advanced calculus there is a $\delta'>0$ (independent of $b$)
such that the solutions with $\alpha<\delta'$ and $|b| \leq B$ are
of type A. By Lemma~10 any $c_n$-orbit has $|b|\leq B$, so for
$\alpha<\delta'$ the $c_n$-orbits are of type A.
\vskip 0.15cm
 By a similar argument as in the proof of Proposition~2 one can
easily show that the type A solutions are generically only $C^0$
at the future SSH (for isolated values of $\alpha$ there are
solutions that go smoothly through the future SSH). In I we showed
that the leading order asymptotic behaviour at the future SSH is
(using $y=x_A-x$)
\begin{equation}
w \sim -1+y, \quad A \sim A_0 - 2\alpha A_0 C^2 y \ln^2(y), \quad
H \sim H_0 - C y \ln(y),
\end{equation}
where $A_0=1-2\alpha \cos^2{H_0}$, $C=\sin(2 H_0)/2 A_0$, and
$H_0$ is a free parameter. Using this expansion one can check that
the curvature is finite as $y \rightarrow 0$ which means that the
type A solutions are examples of naked singularities.
\section*{Acknowledgments} We acknowledge the  hospitality of the Erwin
Schr\"odinger Institute for Mathematical Physics in Vienna, where
part of this paper was produced.  The research of PB was supported
in part by the KBN grant 2 P03B 010 16.
\newpage
\section*{Appendix (local existence theorems)}
In~\cite{bfm} (Proposition~1) Breitelohner, Forg\'acs, and Maison
have derived the following result concerning the behaviour of
solutions of a system of ordinary differential equations near a
singular point (see also~\cite{ren_sch} for a similar result):
\begin{thm2}[BFM]
 Consider a system of first order differential
equations for $n+m$ functions $u=(u_1,...,u_n)$ and
$v=(v_1,...,v_m)$
\begin{equation}\label{bfm}
t\frac{du_i}{dt} = t^{\mu_i} f_i(t,u,v),\qquad t\frac{dv_i}{dt} =
-\lambda_i v_i + t^{\nu_i} g_i(t,u,v),
\end{equation}
where constants $\lambda_i>0$ and integers $\mu_i, \nu_i\geq1$ and
let $C$ be an open subset of $R^n$ such that the functions $f$ and
$g$ are analytic in the neighbourhood of $t=0, u=c, v=0$ for all
$c\in C$. Then there exists an $n$-parameter family of solutions
of the system (\ref{bfm}) such that
\begin{equation}\label{bfm2}
  u_i(t) = c_i + O(t^{\mu_i}), \qquad v_i(t)=O(t^{\nu_i}),
\end{equation}
where $u_i(t)$ and $v_i(t)$ are defined for all $c\in C,
|t|<t_0(c)$ and are analytic in $t$ and $c$.
\end{thm2}
We shall use this theorem to prove existence of local solutions of
equations (21)-(23) near the singular points $\rho=0$ and
$\rho=1$.
\begin{prop}
The equations (21)-(23) admit  a two-parameter family of local
solutions near $\rho=0$
\begin{eqnarray}\label{local}
H(\rho) &=& -\frac{\pi}{2} + c \rho + O(\rho^3),\\
A(\rho) &=& 1 - \alpha c^2 \rho^2 +O(\rho^4),\\
Z(\rho) &=& d \rho +O(\rho^3),
\end{eqnarray}
which are analytic in $c, d$ and $\rho$.
\end{prop}
\noindent {\emph{Proof:}}  Using the variables
\begin{equation}\label{ch1}
  w_1=\frac{H+\pi/2}{\rho}, \quad w_2=H', \quad
  w_3=\frac{1-A}{\rho^2}, \quad w_4=\frac{Z}{\rho}
\end{equation}
we rewrite the equations (21)-(23) as the first order system
\begin{align}\label{s1}
\rho w_1' &= -w_1 +w_2, & \rho w_2' &= 2 w_1- 2 w_2 + \rho^2
h_1, \notag \\
\rho w_3' &= -2 w_3 +2 \alpha w_2^2 + \rho^2 h_2, & \rho w_4'&=
\rho^2 h_3,
\end{align}
where the functions  $h_i$ are analytic near $\rho=0$. Next,
substituting
\begin{align}\label{ch2}
w_1 &= u_1-v_1, &w_2 &= u_1 + 2 v_1, \notag \\
w_3 &= v_2 +\alpha (u_1^2 - 2 v_1^2 - 8 u_1 v_1), &w_4 &= u_2
\end{align}
we  put (\ref{s1}) into the form (\ref{bfm})
\begin{align}\label{fin}
\rho u_1' &= \rho^2 f_1,  & \rho u_2' &= \rho^2 f_2, \notag \\
\rho v_1' &= -3 v_1 +\rho^2 g_1,  &\rho v_2' &= -2 v_2 +\rho^2
g_2,
\end{align} where the functions $f_i, g_i$ are analytic in an open
neighbourhood of  $\rho=0, u_1=c, u_2=d, v_i=0$ for any $c$ and
$d$. Thus, according to the BFM theorem there exists a
two-parameter family of solutions such that
\begin{align}
u_1 &= c   +O(\rho^2), & u_2 &= d+ O(\rho^2),\\
v_1 &= O(\rho^2), & v_2 & =O(\rho^2),
\end{align}
which is equivalent to (66)-(68).
\begin{prop}
The equations (21)-(23) admit  a one-parameter family of local
solutions  near $\rho=1$
\begin{eqnarray}
H(\rho) &=& b (\rho-1) + O((\rho-1)^2),\\
A(\rho) &=& 1-2 \alpha - 2\alpha (1-2 \alpha) b^2 (\rho-1) +
O((\rho-1)^2),\\
Z(\rho) & =& \rho + O((\rho-1)^2)
\end{eqnarray}
which are analytic in $b$ and $\rho$.
 \end{prop}
\noindent {\emph{Proof:}}
 We define the variables
\begin{align}
u &= H', & v_1 & = \frac{H}{\rho-1} - H',\\
v_2 &= \frac{(1-2 \alpha)-A}{\rho-1} - 2\alpha (1-2 \alpha)
{H'}^2, & v_3 &= \frac{Z-1}{\rho-1}-1.
\end{align}
Then, the equations (21)-(23) take the form (using $t=\rho-1$)
\begin{equation}
  t u' = t f, \qquad t v_i' = -v_i + t g_i,
\end{equation}
where the functions $f$ and $g_i$ are analytic in an open
neighourhood of $t=0, u=b, v_i=0$ for any $b>0$. Thus, according
to the BFM theorem there exists a one-parameter family of
solutions such that
\begin{equation}
u(t)= b + O(t), \qquad\,\, v_i(t)=O(t),
\end{equation}
which is equivalent to (75)-(77).


\begin{thebibliography}{10}

\bibitem{bi_wa} P.~Bizo\'n and A. Wasserman, \emph{Self-similar spherically symmetric wave
maps coupled to gravity},
    Phys. Rev. \textbf{D62}, 084031 (2000).

\bibitem{bi} P.~Bizo\'n, \emph{Equivariant self-similar wave maps
from Minkowski spacetime into 3-sphere},
Commun. Math. Phys. \textbf{215}, 45 (2000).

\bibitem{vienna} C. Lechner, J. Thornburg, S. Husa, and P. C. Aichelburg,
\emph{A New Transition between Discrete and Continuous
Self-Similarity in Critical Gravitational Collapse},\\
gr-qc/0112008.


\bibitem{lechner} C. Lechner, Ph. D. Thesis, University of Vienna,
2001.

\bibitem{bfm} P. Breitenlohner, P. Forg\'acs, and D. Maison,
\emph{On static spherically symmetric solutions of the
Einstein-Yang-Mills equations}, Commun. Math. Phys. \textbf{163},
141 (1994).

\bibitem{ren_sch} A. D. Rendall and B. G. Schmidt, \emph{Existence
and properties of spherically symmetric static fluid bodies with a
given equation of state}, Class. Quantum Grav. \textbf{8}, 985
(1991).
\end{thebibliography}
\end{document}